\documentclass[aps,prb,twocolumn,showpacs,amsmath,amssymb]{revtex4}

\usepackage[dvips]{graphicx}
\usepackage{bm}
\usepackage{epsfig}
\usepackage[german,english]{babel}

\begin{document}

\title{Ferromagnetism in multi-band Kondo lattice model}
\author{A.Sharma}
\email{anand@physik.hu-berlin.de}

\author{W.Nolting}

\affiliation{Institut f$\ddot{u}$r Physik, Humboldt-Universit$\ddot{a}$t zu Berlin, Newtonstr.15, 12489, Berlin, Germany.}

\date{\today}

\begin{abstract}
The ferromagnetic spin exchange interaction between the itinerant electrons and localized moments on a periodic lattice, studied within the so-called Kondo lattice model (KLM), is considered for multiband situation where the hopping integral is a matrix, in general. The modified RKKY theory, wherein one can map such a model onto an effective Heisenberg-like system, is extended to a multi-band case with finite bandwidth and hybridization on a simple cubic lattice. As an input for the evaluation of the effective exchange integrals, one requires the multi-band electronic self energy which is taken from an earlier proposed ansatz. Using the above procedure, we determine the magnetic properties of the system like Curie temperature while calculating the chemical potential and magnetization within a self consistent scheme for various values of system parameters. The results are discussed in detail and the model is motivated in order to study the electronic, transport and magnetic properties of real materials like GdN.
\end{abstract}


\maketitle

\section{Introduction}

\indent The original \textit{Kondo model}~\cite{kondo} with anti-ferromagnetic spin-exchange interaction between a single impurity spin in a non-magnetic background and the itinerant electrons of the host metal was used by Jun Kondo to explain the unusual temperature behaviour of the resistivity of the system. Its periodic extension with ferromagnetic exchange interaction between a system of localized spins and a band of itinerant electrons is in the literature often referred to as the \textit{ferromagnetic Kondo lattice model}~\cite{furu,dago} or \textit{s-d}~\cite{kasu,yosi} or \textit{s-f} model~\cite{nol79}. For the sake of uniformity, we ascribe it as the \textit{Kondo lattice model} (KLM). Both the anti-ferromagnetic as well as the ferromagnetic alignment of itinerant and localized spins exhibits remarkable differences in the physical properties of various real materials and has been a subject of intense theoretical studies in the past. \\
\indent For instance, the magnetic semiconductors (prototypes being the europium chalcogenides : EuX : X=O,S,Se,Te)~\cite{nagaev,wachter,mauger} are known to have ferromagnetic exchange coupling and demonstrate a spectacular temperature dependence of the band states. The redshift of the optical absorption edge in these materials upon cooling from T=T$_{\textrm{c}}$ to T=0 K is due to a corresponding shift of the lower conduction band edge~\cite{wachter,mueller}. A great deal of focus has been concentrated on studying the diluted magnetic semiconductors with anti-ferromagnetic~\cite{dietl} and ferromangetic~\cite{tang,singh} exchange interaction with the purpose of achieving practical spintronics~\cite{wolf,zutic} applications. Apart from magnetic semiconductors, the local moment metals like Gd are known to have a ferromagnetic exchange~\cite{rexeyenol}. But the exchange induced correlation and the temperature dependent quasi-particle effects~\cite{santos} have lead to complex and hence controversial photoemission data~\cite{lizhang,dondownol}. Other materials like the manganese oxides (manganites) having pervoskite structures (the prototype being  A$_{1-\textrm{x}}$B$_{\textrm{x}}$MnO$_{3}$  where A=La,Pr,Nd and B=Sr,Ca,Ba,Pb) also have a strong ferromagnetic exchange interaction. They have a remarkable property called colossal magnetoresistance (CMR)~\cite{jin,ramirez} which enables them to dramatically change their electrical resistance in the presence of a magnetic field. Many theoretical models have been proposed in order to explain the existence of these effects. The earlier theoretical ideas were based up on the double exchange model~\cite{doublexchange} which can be understood as one of the limiting case of Kondo lattice model (i.e. limit of strong Hunds coupling). Though recent theories~\cite{stiernol} have provided a step forward but its complete understanding is far from being explained by any current physical theories. As compared to the aforementioned compounds, the heavy fermion systems~\cite{loeh} (mostly Ce compounds) are known to have anti-parallel alignment of the conduction electron and localized spins. They have been rigorously studied because of their extraordinary physical properties~\cite{doniach}. \\
\indent In the above examples, the kinetic energy of the itinerant electrons is usually described within tight binding dispersion of a single non-degenerate band i.e., single orbital atom per unit cell. But it is well known that the single band calculations are certainly not sufficient in order to have a complete understanding of unusual phenomenon in real materials~\cite{popescu}. One has to take into account the intra and inter- band interactions as well. \\
\indent The multi-band models are also of growing interest for exhibiting a wide range of phenomena like novel electronic phases, magnetism and superconductivity~\cite{nagao}. For instance, it was found out using a two-band Hubbard~\cite{hub} model that there was a possibility of existence of ferromagnetism around half-filling~\cite{penc} in contrast to antiferromagnetism in single band model~\cite{nolbor}. The numerical studies~\cite{wang} on ground state properties of multi-band periodic Anderson~\cite{pam} model revealed the minor role played by the competition between RKKY~\cite{rkky} and Kondo interactions~\cite{kondo} again in contrast to the single band case~\cite{meyer}. This motivates us to understand the physics behind the interplay between the kinetic and potential energy of multi-band Kondo lattice model and extend it for studying the electronic, transport and magnetic properties of real materials like GdN. \\
\indent This paper is organized as follows. In the following section we develop our theoretical multi-band model Hamiltonian describing the physics due to the intra-atomic exchange interaction between the two sub-systems, i.e., itinerant electrons and localized spins, on a periodic lattice. In section~\ref{sec:elesubsys}, we consider only the electronic part of the system while treating the magnetic part within molecular-field theory. Using an earlier proposed ansatz for multi-band self-energy~\cite{sharmagdn}, we evaluate the electronic properties of interest like the density of states and band occupation number. In section~\ref{sec:magsubsys}, we develop the modified RKKY theory~\cite{nolrexmat} wherein we integrate out the charge degrees of freedom of the itinerant electrons thereby mapping the multi-band Kondo lattice Hamiltonian onto an effective Heisenberg- like spin Hamiltonian. In section~\ref{sec:resdis}, using the procedure described in earlier section we determine the magnetic properties of the system like Curie temperature while calculating the chemical potential and magnetization within a self consistent scheme. We discuss the results as obtained for various values of system parameters. In section~\ref{sec:sumcon}, we summarize and conclude our findings. 

\section{Model and Methods}
\label{sec:theomod}

\subsection{Electronic sub-system}
\label{sec:elesubsys}

In this section, we present a brief description of the theoretical model used in our calculations. The details of the many body analysis along with a model calculation and limiting cases are explained elsewhere~\cite{sharmagdn}. The multi-band Kondo lattice model (KLM) Hamiltonian mainly consists of two parts ; 
\begin{equation}\label{eq:Ham}
H=H_{kin}+H_{int}
\end{equation}
where \\
\begin{equation}\label{eq:Ho}
H_{kin}=\sum_{ij\alpha\beta\sigma}
T_{ij}^{\alpha\beta}c_{i\alpha\sigma}^{\dagger}c_{j\beta\sigma}
\end{equation}
and
\begin{equation}\label{eq:Hint}
H_{int}= - \frac{J}{2} \sum_{\substack{i\alpha \\ \sigma^{\prime}\sigma}} (\textbf{S}_{i} \cdot \bm{\sigma})_{\sigma^{\prime}\sigma} c_{i\alpha\sigma^{\prime}}^{\dagger}c_{i\alpha\sigma}
\end{equation}
\indent $H_{kin}$ denotes the kinetic energy of the itinerant electrons with $T_{ij}^{\alpha\beta}$ being the hopping term which is connected by Fourier transformation to the free Bloch energies $\epsilon^{\alpha\beta}(\bold{k})$
\begin{equation}\label{eq:Hop}
\indent T_{ij}^{\alpha\beta}=\frac{1}{N} \sum_{\bold{k}}\epsilon^{\alpha\beta}(\bold{k}) \hspace{0.2cm} e^{-i{\bold{k}} \cdot (R_{i}-R_{j})}
\end{equation}
while $c_{i\alpha\sigma}^{\dagger}$ and $c_{i\alpha\sigma}$ are the fermionic creation and annihilation operators, respectively, at lattice site $R_{i}$. The latin letters (i,j,...) symbolize the crystal lattice indices while the band indices are depicted in Greek letters ($\alpha$,$\beta$,..) and the spin is denoted as ${\sigma}(={\uparrow},{\downarrow})$. \\
\indent $H_{int}$ is an intra-atomic exchange interaction term i.e., a local interaction between electron spin $\bm{\sigma}$ and local moment spin $S_{i}$. Using second quantization for electron spin ($n_{i\alpha\sigma} = c_{i\alpha\sigma}^{\dagger} c_{i\alpha\sigma}$), the interaction term is further being further split into two subterms. 
\begin{equation}
H_{int}=-\frac{J}{2}\sum_{i\alpha\sigma}(z_{\sigma}S_{i}^{z}c_{i\alpha\sigma}^{\dagger}c_{i\alpha\sigma}+S_{i}^{\sigma}c_{i\alpha-\sigma}^{\dagger}c_{i\alpha\sigma})
\end{equation}
\indent The first describes the Ising type interaction between the z-component of the localized and itinerant carrier spins while the other comprises spin exchange processes which are responsible for many of the KLM properties. J is the exchange coupling strength which we assume to be $\textbf{k}$-independent and $S_{i}^{\sigma}$ refers to the localized spin at site $R_{i}$
\begin{equation}\label{eq:Loc-spin}
\indent S_{i}^{\sigma}=S_{i}^{x}+iz_{\sigma}S_{i}^{y} \hspace{0.2cm}; \hspace{0.2cm} z_{\uparrow}=+1, z_{\downarrow}=-1
\end{equation}
\indent The Hamiltonian in eq.~\eqref{eq:Ham} provokes a nontrivial many body problem that cannot be solved exactly.
Using the equation of motion method for the double-time retarded Green function~\cite{zubarev}
\begin{equation}\label{eq:Green-fun}
G_{lm\sigma}^{\mu\nu}(E)=\langle\langle c_{l\mu\sigma};c_{m\nu\sigma}^\dagger\rangle\rangle_{E}
\end{equation}
where l,m and $\mu$,$\nu$ are the lattice and band indices respectively, we obtain higher order Green functions which prevent the direct solution. Approximations must be considered. But a rather formal solution can be stated as
\begin{equation}\label{eq:Green-matrix}
\widehat{G}_{\textbf{k}\sigma}(E)=[{(E+i0^{+})\widehat{I}-\widehat{\epsilon}(\bold{k})-\widehat{\Sigma}_{\textbf{k}\sigma}(E)}]^{-1}
\end{equation}
where for simplicity we exclude the band indices by representing the terms in a generalized matrix form on symbolizing a hat over it
\begin{equation}\label{eq:Green-Fourier}
\widehat{G}_{lm\sigma}(E)=\frac{1}{N}\sum_{\textbf{k}}\widehat{G}_{\textbf{k}\sigma}(E) \hspace{0.2cm} e^{-i{\textbf{k}} \cdot (R_{l}-R_{m})}
\end{equation}
\indent The terms in eq.~\eqref{eq:Green-matrix} are explained as follows : $\hat{I}$ is an identity matrix. 0$^{+}$ is small imaginary part and $\hat{\epsilon}(\bold{k})$ is a hopping matrix with the diagonal terms of the matrix exemplifying the intra-band hopping and the off-diagonal terms denoting the inter-band hopping. The self energy, $\widehat{\Sigma}_{\textbf{k} \sigma}$(E), containing all the influences of the different interactions being of fundamental
importance, can be understood using site representation :
\begin{equation}\label{eq:self-energy}
\langle\langle[H_{int},c_{l\mu\sigma}]_{-};c_{m\nu\sigma}^{\dagger}
\rangle\rangle=\sum_{p\gamma}\Sigma_{lp\sigma}^{\mu\gamma}(E)G_{pm\sigma}^{\gamma\nu}(E)
\end{equation}
\indent Now we are left with a problem of finding a multi-band self energy ansatz in order to compute the Green function matrix and thereby calculate the physical quantities of interest like the quasi-particle spectral density (SD)
\begin{equation}\label{eq:SD}
S_{\textbf{k}\sigma}(E)=-\frac{1}{\pi}Im Tr(\widehat{G}_{\textbf{k}\sigma}(E))
\end{equation}
and the quasi-particle density of states (Q-DOS)
\begin{equation}\label{eq:DOS}
\rho_{\sigma}(E)=\frac{1}{N\hbar}\sum_{\bold{k}}S_{\textbf{k}\sigma}(E)
\end{equation}
which would yield the band occupation number
\begin{equation}
n = \sum_{\sigma} n_{\sigma} = \int_{-\infty}^{\infty} dE \hspace*{0.1cm} f_{-}(E) \hspace*{0.1cm} \rho_{\sigma}(E)
\end{equation}
where 
\begin{equation}
f_{-}(E) = \frac{1}{e^{\frac{E-\mu}{k_{B}T}}+1} \nonumber
\end{equation}
is the Fermi function and $\mu$ being the chemical potential or the Fermi edge.\\
\indent According to our many body theoretical analysis~\cite{sharmagdn}, we utilize the multi-band interpolating self energy ansatz (ISA) which is well defined in the low carrier density regime~\cite{sharmaeus}, for all coupling strengths and satisfying one limiting case of the model namely that of ferromagnetically saturated semiconductor. The ansatz is given as:
\begin{subequations}
\begin{multline}\label{eq:Mult-SE}
\widehat{\Sigma}_{\sigma}(E)=\frac{J}{2}M_{-\sigma}\widehat{I}+\frac{J^{2}}{4}a_{-\sigma}\widehat{G}_{-\sigma}\bigg(E+\frac{J}{2}M_{-\sigma}\bigg)\\
\left[\widehat{I}-\frac{J}{2}\widehat{G}_{-\sigma}\bigg(E+\frac{J}{2}M_{-\sigma}\bigg)\right]^{-1}
\end{multline}
where
\begin{equation}
M_{\sigma}=z_{\sigma} \langle S^{z} \rangle;
\hspace{0.2cm}a_{\sigma}=S(S+1)+M_{\sigma}(M_{\sigma}+1).
\end{equation}
\end{subequations}
and the bare Green function matrix is defined as :
\begin{equation}\label{eq:Eff-Green}
\widehat{G}_{\sigma}(E)=\frac{1}{N}\sum_{\textbf{k}}[(E+i0^{+})\widehat{I}-\widehat{\epsilon}(\bold{k})]^{-1} \nonumber\\
\end{equation}
\indent The first term in Eq.~(\ref{eq:Mult-SE}), which is exact in the weak coupling limit, represents an induced \textit{Stoner} \textit{splitting} of the energy band proportional to the \textit{f} spin magnetization $\langle S^{z} \rangle$. The second term is dominated by the consequences of spin exchange processes between itinerant electrons and localized \textit{f} moments. \\
\indent As seen in Eq.~(\ref{eq:Mult-SE}), we are interested only in the local self energy
\begin{equation}\label{eq:SE-k-ind}
\widehat{\Sigma}_{\sigma}(E)=\frac{1}{N}\sum_{\bold{k}}\widehat{\Sigma}_{\bold{k}\sigma}(E)\nonumber\\
\end{equation}
while the wave-vector dependence of the self energy is mainly due to the magnon energies appearing at finite temperature. In order to evaluate only the itinerant electron subsystem, we can neglect this wave-vector dependence. The localized magnetization $\langle S^{z} \rangle$ can then be considered as an external parameter being responsible for the induced temperature dependence of the band states. Thus in a non-self consistent way, it is possible to determine the influence of inter- band exchange on the conduction band states so as to study the electronic correlations effects~\cite{sharmagdn,sharmaeus}. But in order to study the effect of itinerant electron subsystem on the localized subsystem and vice versa, we need to calculate the magnetization within a self-consistent manner as shown in the following section.

\subsection{Magnetic sub-system}
\label{sec:magsubsys}

\indent In Section~\ref{sec:elesubsys}, we did not consider a direct exchange interaction between the localized \textit{f} spins. But if one is interested in determining the magnetic properties of multi-band Kondo lattice model then both the sub-systems (localized as well as itinerant) should be solved within a self consistent scheme. 
\begin{figure}[!h]
\centering
\includegraphics[height=4cm,width=7cm]{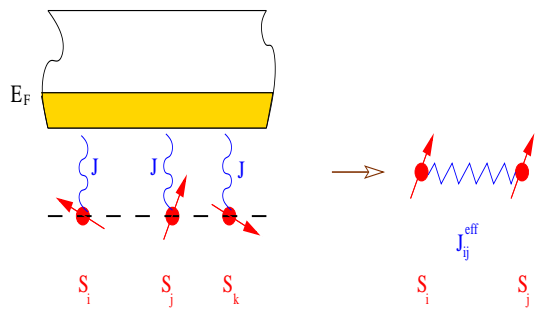}
\caption{(Color online) An effective indirect exchange, $J^{\textrm{eff}}_{\textrm{ij}}$, between localized \textit{f} spins (red arrows) mediated by intra-atomic exchange, $J$, due to itinerant electrons. E$_{F}$ denotes the Fermi edge. \label{fig:indirx}}
\end{figure}
Therefore, we would like to take into account an effective indirect coupling, J$^{\textrm{eff}}_{\textrm{ij}}$, between the localized \textit{f} spins and the itinerant electrons within the so-called modified RKKY~\cite{nolrexmat} formalism as shown in Figure~\ref{fig:indirx}. \\
\indent Consider the multi-band Kondo lattice Hamiltonian, i.e. Eq.~(\ref{eq:Ham}), which can be written in the following equivalent form,
\begin{eqnarray}
H \nonumber
& = & H_{kin} + H_{int} \nonumber \\
& = & \sum_{\textbf{k}\alpha\beta\sigma} \epsilon^{\alpha\beta}(\textbf{k}) c_{\textbf{k}\alpha\sigma}^{\dagger}c_{\textbf{k}\beta\sigma} \nonumber \\ 
& - & \frac{J}{2N} \sum_{\substack{i\alpha \\ \sigma^{\prime}\sigma}} \sum_{\textbf{k}\textbf{q}} e^{-i\textbf{q} \cdot \textbf{R}_{i}} (\textbf{S}_{i} \cdot \bm{\sigma})_{\sigma^{\prime}\sigma} c_{\textbf{k+q}\alpha\sigma^{\prime}}^{\dagger}c_{\textbf{k}\alpha\sigma} \nonumber
\end{eqnarray}
where all the terminologies remain the same as explained in Section~\ref{sec:elesubsys}. The components of the band electron spin operator $\bm{\sigma}$ are the Pauli spin matrices.\\
\indent The main idea of the modified RKKY theory is to transform the above Kondo-like exchange Hamiltonian of the conduction electrons into an effective Heisenberg-like spin exchange Hamiltonian of the \textit{f} spins by averaging $H_{int}$ in the subspace of the conduction electrons \\ ( $\langle$ \hspace*{0.2cm} $\rangle$ ):
\begin{eqnarray}\label{eq:av}
\langle H_{int} \rangle \nonumber
& = & H_{f} \nonumber \\
& = & - \frac{J}{2N} \sum_{\substack{i\alpha \\ \sigma^{\prime}\sigma}}\sum_{\textbf{k}\textbf{q}} e^{-i\textbf{q} \cdot \textbf{R}_{i}} (\textbf{S}_{i} \cdot \bm{\sigma})_{\sigma^{\prime}\sigma} \langle c_{\textbf{k+q}\alpha\sigma^{\prime}}^{\dagger}c_{\textbf{k}\alpha\sigma} \rangle \nonumber \\
\end{eqnarray}
\indent This is achieved by closely following the treatment as given in Ref.~\onlinecite{nolrexmat}. For averaging procedure the \textit{f} spin operators are to be considered as \textit{c} numbers. The expectation values $\langle$ \hspace*{0.2cm} $\rangle$ in Eq.~(\ref{eq:av}) may still have operator properties in the \textit{f} spin subspace and therefore do not vanish for $\textbf{q} \neq 0$ and $\sigma \neq \sigma^{\prime}$. We would like to obtain $\langle$ \hspace*{0.2cm} $\rangle$ via the spectral theorem with the help of appropriate Green function as given below :
\begin{equation}\label{eq:gfnw}
G_{\textbf{k},\textbf{k+q}}^{\alpha\beta\sigma\sigma^{\prime}}(E) = \langle\langle c_{\textbf{k}\alpha\sigma};c_{\textbf{k+q}\beta\sigma^{\prime}}^{\dagger} \rangle\rangle_{E}
\end{equation}
\indent Its equation of motion can be obtained in the usual way~\cite{zubarev} and is given as
\begin{align}\label{eq:eomgf}
E G_{\textbf{k},\textbf{k+q}}^{\alpha\beta\sigma\sigma^{\prime}}(E) = \delta_{\textbf{k},\textbf{k+q}}\delta_{\alpha\beta}\delta_{\sigma\sigma^{\prime}} 
+ \sum_{\gamma} \epsilon^{\alpha\gamma}(\textbf{k}) G_{\textbf{k},\textbf{k+q}}^{\gamma\beta\sigma\sigma^{\prime}}(E) \hspace*{2.0cm} \nonumber \\
-\frac{J}{2N} \sum_{i\textbf{k}^{\prime}\sigma^{\prime\prime}} e^{-i(\textbf{k}-\textbf{k}^{\prime}) \cdot \textbf{R}_{i}} (\textbf{S}_{i} \cdot \bm{\sigma})_{\sigma\sigma^{\prime\prime}} G_{\textbf{k}^{\prime},\textbf{k+q}}^{\alpha\beta\sigma^{\prime\prime}\sigma^{\prime}}(E) \hspace*{2.0cm} 
\end{align}
\indent The above equation can be iterated up to any desired accuracy producing spin products of the type :\\
\hspace*{0.2cm} $(\textbf{S}_{i} \cdot \bm{\sigma})_{\sigma\sigma^{\prime\prime}}$, \hspace*{0.2cm} $(\textbf{S}_{i} \cdot \bm{\sigma})_{\sigma^{\prime\prime}\sigma^{\prime\prime\prime}}$, \hspace*{0.2cm} $(\textbf{S}_{i} \cdot \bm{\sigma})_{\sigma^{\prime\prime\prime}\sigma^{\prime\prime\prime\prime}}$ \\
\indent On excluding the band indices in Eq.~(\ref{eq:eomgf}) by representing the terms in a generalized matrix form on symbolizing a hat over it we get
\begin{align}\label{eq:eomgfm}
E \widehat{G}_{\textbf{k},\textbf{k+q}}^{\sigma\sigma^{\prime}}(E) = \delta_{\textbf{k},\textbf{k+q}} \delta_{\sigma\sigma^{\prime}} \widehat{I} + \widehat{\epsilon}(\textbf{k}) \widehat{G}_{\textbf{k},\textbf{k+q}}^{\sigma\sigma^{\prime}}(E) \hspace*{4.0cm} \nonumber \\
-\frac{J}{2N} \sum_{i\textbf{k}^{\prime}\sigma^{\prime\prime}} e^{-i(\textbf{k}-\textbf{k}^{\prime}) \cdot \textbf{R}_{i}} (\textbf{S}_{i} \cdot \bm{\sigma})_{\sigma\sigma^{\prime\prime}} \widehat{G}_{\textbf{k}^{\prime},\textbf{k+q}}^{\sigma^{\prime\prime}\sigma^{\prime}}(E) \hspace*{2.5cm}
\end{align}
\indent Rearranging the terms in Eq.~(\ref{eq:eomgfm}) yields 
\begin{align}\label{eq:eomfa}
\big[E\widehat{I} - \widehat{\epsilon}(\textbf{k})\big] \widehat{G}_{\textbf{k},\textbf{k+q}}^{\sigma\sigma^{\prime}}(E) = \delta_{\textbf{k},\textbf{k+q}}\delta_{\sigma\sigma^{\prime}} \widehat{I} \hspace*{5.0cm} \nonumber \\
-\frac{J}{2N} \sum_{i\textbf{k}^{\prime}\sigma^{\prime\prime}} e^{-i(\textbf{k}-\textbf{k}^{\prime}) \cdot \textbf{R}_{i}} (\textbf{S}_{i} \cdot \bm{\sigma})_{\sigma\sigma^{\prime\prime}} \widehat{G}_{\textbf{k}^{\prime},\textbf{k+q}}^{\sigma^{\prime\prime}\sigma^{\prime}}(E) \hspace*{2.0cm}
\end{align}
\indent For symmetry reasons, we write down the equation of motion for $G_{\textbf{k},\textbf{k+q}}^{\alpha\beta\sigma\sigma^{\prime}}(E)$ in an alternative way where the second operator , $c_{\textbf{k+q}\beta\sigma^{\prime}}^{\dagger}$ , in Eq.~(\ref{eq:gfnw}) is the "active" operator.: 
\begin{align}
E \widehat{G}_{\textbf{k},\textbf{k+q}}^{\sigma\sigma^{\prime}}(E)
= \delta_{\textbf{k},\textbf{k+q}}\delta_{\sigma\sigma^{\prime}} \widehat{I} 
+ \widehat{G}_{\textbf{k},\textbf{k+q}}^{\sigma\sigma^{\prime}}(E) \widehat{\epsilon}(\textbf{k+q}) \hspace*{3.0cm} \nonumber \\
- \frac{J}{2N} \sum_{i\textbf{k}^{\prime}\sigma^{\prime\prime}} e^{-i\big(\textbf{k}^{\prime}-(\textbf{k+q})\big) \cdot \textbf{R}_{i}} (\textbf{S}_{i} \cdot \bm{\sigma})_{\sigma^{\prime\prime}\sigma^{\prime}} \widehat{G}_{\textbf{k},\textbf{k}^{\prime}}^{\sigma\sigma^{\prime\prime}}(E) \hspace*{2.0cm}
\end{align}
and again upon rearranging the terms in the above equation we get
\begin{align}\label{eq:eomsa}
\widehat{G}_{\textbf{k},\textbf{k+q}}^{\sigma\sigma^{\prime}}(E) \big[E\widehat{I} - \widehat{\epsilon}(\textbf{k+q})\big] = \delta_{\textbf{k},\textbf{k+q}}\delta_{\sigma\sigma^{\prime}} \widehat{I} \hspace*{5.0cm} \nonumber \\
-\frac{J}{2N} \sum_{i\textbf{k}^{\prime}\sigma^{\prime\prime}} e^{-i\big(\textbf{k}^{\prime}-(\textbf{k+q})\big) \cdot \textbf{R}_{i}} (\textbf{S}_{i} \cdot \bm{\sigma})_{\sigma^{\prime\prime}\sigma^{\prime}} \widehat{G}_{\textbf{k},\textbf{k}^{\prime}}^{\sigma\sigma^{\prime\prime}}(E) \hspace*{3.0cm}
\end{align}
\indent Now let us define the following Green function of the "free" electron system
\begin{equation}\label{eq:fgf1}
\big[E\widehat{I} - \widehat{\epsilon}(\textbf{k})\big] = \big(\widehat{G}_{\textbf{k}}^{(0)}(E)\big)^{-1}
\end{equation}
\begin{equation}\label{eq:fgf2}
\big[E\widehat{I} - \widehat{\epsilon}(\textbf{k+q})\big] = \big(\widehat{G}_{\textbf{k+q}}^{(0)}(E)\big)^{-1}
\end{equation}
\indent Subsituting Eq.~(\ref{eq:fgf1}) in Eq.~(\ref{eq:eomfa}) gives
\begin{align}\label{eq:eomefa}
\big(\widehat{G}_{\textbf{k}}^{(0)}(E)\big)^{-1} \widehat{G}_{\textbf{k},\textbf{k+q}}^{\sigma\sigma^{\prime}}(E) \nonumber
= \delta_{\textbf{k},\textbf{k+q}}\delta_{\sigma\sigma^{\prime}} \widehat{I} \hspace*{5.0cm} \nonumber \\
-\frac{J}{2N} \sum_{i\textbf{k}^{\prime}\sigma^{\prime\prime}} e^{-i(\textbf{k}-\textbf{k}^{\prime}) \cdot \textbf{R}_{i}} (\textbf{S}_{i} \cdot \bm{\sigma})_{\sigma\sigma^{\prime\prime}} \widehat{G}_{\textbf{k}^{\prime},\textbf{k+q}}^{\sigma^{\prime\prime}\sigma^{\prime}}(E) \hspace*{2.0cm}
\end{align}
while substituting Eq.~(\ref{eq:fgf2}) in Eq.~(\ref{eq:eomsa}) yields
\begin{align}\label{eq:eomesa}
\widehat{G}_{\textbf{k},\textbf{k+q}}^{\sigma\sigma^{\prime}}(E) \big(\widehat{G}_{\textbf{k+q}}^{(0)}(E)\big)^{-1} \nonumber
= \delta_{\textbf{k},\textbf{k+q}}\delta_{\sigma\sigma^{\prime}} \widehat{I} \hspace*{5.0cm}\nonumber \\
-\frac{J}{2N} \sum_{i\textbf{k}^{\prime}\sigma^{\prime\prime}} e^{-i\big(\textbf{k}^{\prime}-(\textbf{k+q})\big) \cdot \textbf{R}_{i}} (\textbf{S}_{i} \cdot \bm{\sigma})_{\sigma^{\prime\prime}\sigma^{\prime}} \widehat{G}_{\textbf{k},\textbf{k}^{\prime}}^{\sigma\sigma^{\prime\prime}}(E) \hspace*{2.0cm}
\end{align}
\indent Now upon multiplying $\widehat{G}_{\textbf{k}}^{(0)}(E)$ from left to Eq.~(\ref{eq:eomefa}) we get
\begin{align}\label{eq:gf1}
\widehat{G}_{\textbf{k},\textbf{k+q}}^{\sigma\sigma^{\prime}}(E)
= \delta_{\textbf{k},\textbf{k+q}}\delta_{\sigma\sigma^{\prime}} \widehat{G}_{\textbf{k}}^{(0)}(E) \hspace*{3.5cm}\nonumber \\
-\frac{J}{2N} \sum_{i\textbf{k}^{\prime}\sigma^{\prime\prime}} e^{-i(\textbf{k}-\textbf{k}^{\prime}) \cdot \textbf{R}_{i}} (\textbf{S}_{i} \cdot \bm{\sigma})_{\sigma\sigma^{\prime\prime}} \widehat{G}_{\textbf{k}}^{(0)}(E) \widehat{G}_{\textbf{k}^{\prime},\textbf{k+q}}^{\sigma^{\prime\prime}\sigma^{\prime}}(E) 
\end{align}
\indent And on multiplying $\widehat{G}_{\textbf{k+q}}^{(0)}(E)$ from right to Eq.~(\ref{eq:eomesa}) we obtain
\begin{align}\label{eq:gf2}
\widehat{G}_{\textbf{k},\textbf{k+q}}^{\sigma\sigma^{\prime}}(E)
= \delta_{\textbf{k},\textbf{k+q}}\delta_{\sigma\sigma^{\prime}} \widehat{G}_{\textbf{k+q}}^{(0)}(E) \hspace*{6.0cm}\nonumber\\
- \frac{J}{2N} \sum_{i\textbf{k}^{\prime}\sigma^{\prime\prime}} e^{-i\big(\textbf{k}^{\prime}-(\textbf{k+q})\big) \cdot \textbf{R}_{i}} (\textbf{S}_{i} \cdot \bm{\sigma})_{\sigma^{\prime\prime}\sigma^{\prime}} \widehat{G}_{\textbf{k},\textbf{k}^{\prime}}^{\sigma\sigma^{\prime\prime}}(E) \widehat{G}_{\textbf{k+q}}^{(0)}(E)\hspace*{2.0cm}
\end{align}
\indent Let us make the following crucial first order approximations for the Green functions 
\begin{equation}\label{eq:app1}
\widehat{G}_{\textbf{k},\textbf{k}^{\prime}}^{\sigma\sigma^{\prime\prime}}(E) \approx \delta_{\sigma\sigma^{\prime\prime}}\delta_{\textbf{k},\textbf{k}^{\prime}} \widehat{G}_{\textbf{k}\sigma}(E)
\end{equation}
\begin{equation}\label{eq:app2}
\widehat{G}_{\textbf{k}^{\prime},\textbf{k+q}}^{\sigma^{\prime\prime}\sigma^{\prime}}(E) \approx \delta_{\sigma^{\prime\prime}\sigma^{\prime}}\delta_{\textbf{k}^{\prime},\textbf{k+q}} \widehat{G}_{\textbf{k+q}\sigma^{\prime}}(E)
\end{equation}
where
\begin{equation}\label{eq:igf1}
\widehat{G}_{\textbf{k}\sigma}(E) = \big[E\widehat{I} - \widehat{\epsilon}(\textbf{k}) - \widehat{\Sigma}_{\sigma}(E)\big]^{-1}
\end{equation}
\begin{equation}\label{eq:igf2}
\widehat{G}_{\textbf{k+q}\sigma^{\prime}}(E) = \big[E\widehat{I} - \widehat{\epsilon}(\textbf{k+q}) - \widehat{\Sigma}_{\sigma^{\prime}}(E)\big]^{-1} 
\end{equation}
The renormalization by the interacting Green functions as performed in Eq.~(\ref{eq:app1}) and Eq.~(\ref{eq:app2}) should be a sensible approximation since it is observed that if those interacting Green functions 
are replaced by the free Green functions, Eq.~(\ref{eq:fgf1}) and Eq.~(\ref{eq:fgf2}) respectively, then it leads to the correct low-$J$ (i.e. RKKY) behaviour. On substituting Eq.~(\ref{eq:app2}) in Eq.~(\ref{eq:gf1}) we obtain 
\begin{align}\label{eq:gf3}
\widehat{G}_{\textbf{k},\textbf{k+q}}^{\sigma\sigma^{\prime}}(E) \nonumber
= \delta_{\textbf{q},0}\delta_{\sigma\sigma^{\prime}} \widehat{G}_{\textbf{k}}^{(0)}(E) \hspace*{3.5cm}\nonumber \\
- \frac{J}{2N} \sum_{i} e^{i\textbf{q} \cdot \textbf{R}_{i}} (\textbf{S}_{i} \cdot \bm{\sigma})_{\sigma\sigma^{\prime}} \widehat{G}_{\textbf{k}}^{(0)}(E) \widehat{G}_{\textbf{k+q}\sigma^{\prime}}(E)
\end{align}
while substituting Eq.~(\ref{eq:app1}) in Eq.~(\ref{eq:gf2}) gives 
\begin{align}\label{eq:gf4}
\widehat{G}_{\textbf{k},\textbf{k+q}}^{\sigma\sigma^{\prime}}(E) \nonumber
= \delta_{\textbf{q},0}\delta_{\sigma\sigma^{\prime}} \widehat{G}_{\textbf{k}}^{(0)}(E) \hspace*{3.5cm}\nonumber \\ 
- \frac{J}{2N} \sum_{i} e^{i\textbf{q} \cdot \textbf{R}_{i}} (\textbf{S}_{i} \cdot \bm{\sigma})_{\sigma\sigma^{\prime}} \widehat{G}_{\textbf{k}\sigma}(E) \widehat{G}_{\textbf{k+q}}^{(0)}(E)
\end{align}
\indent On adding Eq.~(\ref{eq:gf3}) and Eq.~(\ref{eq:gf4}) we get
\begin{align}\label{eq:gffin}
\widehat{G}_{\textbf{k},\textbf{k+q}}^{\sigma\sigma^{\prime}}(E) \nonumber
= \delta_{\textbf{q},0}\delta_{\sigma\sigma^{\prime}} \widehat{G}_{\textbf{k}}^{(0)}(E) \hspace*{3.5cm}\nonumber\\
- \frac{J}{4N} \sum_{i} e^{i\textbf{q} \cdot \textbf{R}_{i}} (\textbf{S}_{i} \cdot \bm{\sigma})_{\sigma\sigma^{\prime}} \widehat{A}_{\textbf{k},\textbf{k+q}}^{\sigma\sigma^{\prime}}(E)
\end{align}
where
\begin{equation}\label{eq:akkpq}
\widehat{A}_{\textbf{k},\textbf{k+q}}^{\sigma\sigma^{\prime}}(E) = \big(\widehat{G}_{\textbf{k}}^{(0)}(E) \widehat{G}_{\textbf{k+q}\sigma^{\prime}}(E) + \widehat{G}_{\textbf{k}\sigma}(E) \widehat{G}_{\textbf{k+q}}^{(0)}(E) \big)
\end{equation}
\indent For the effective spin Hamiltonian in Eq.~(\ref{eq:av}) we need the expectation value $\langle c_{\textbf{k+q}\alpha\sigma^{\prime}}^{\dagger}c_{\textbf{k}\alpha\sigma} \rangle$, which we express in terms of the trace of imaginary part of the Green function Eq.~(\ref{eq:gffin}) by exploiting the spectral theorem~\cite{zubarev} :
\begin{align}\label{eq:spthm}
\frac{1}{N} \sum_{\textbf{k}}\langle c_{\textbf{k+q}\alpha\sigma^{\prime}}^{\dagger}c_{\textbf{k}\alpha\sigma}  \rangle \hspace*{5.5cm}\nonumber \\
= -\frac{1}{\pi N} Im Tr \int_{-\infty}^{\infty} dE f_{-}(E) \sum_{\textbf{k}} \widehat{G}_{\textbf{k},\textbf{k+q}}^{\sigma\sigma^{\prime}}(E) \hspace*{1.3cm}\nonumber \\
= \delta_{\textbf{q},0}\delta_{\sigma\sigma^{\prime}} \bigg( \frac{-1}{\pi N}\bigg) Im Tr \int_{-\infty}^{\infty} dE f_{-}(E) \sum_{\textbf{k}} \widehat{G}_{\textbf{k}}^{(0)}(E) \nonumber \\
+ \frac{J}{4\pi N^{2}} \sum_{i} \bigg[ e^{i\textbf{q} \cdot \textbf{R}_{i}} (\textbf{S}_{i} \cdot \bm{\sigma})_{\sigma\sigma^{\prime}} \hspace*{0.25cm} * \hspace*{2.5cm}\nonumber \\
* \hspace*{0.25cm} Im Tr \int_{-\infty}^{\infty} dE f_{-}(E) \sum_{\textbf{k}} \widehat{A}_{\textbf{k},\textbf{k+q}}^{\sigma\sigma^{\prime}}(E) \bigg]
\end{align}
\indent On substituing Eq.~(\ref{eq:spthm}) in Eq.~(\ref{eq:av}) we get \\ \\
$H_{f}$ =
\begin{align}
\frac{J}{2\pi N} \sum_{i\sigma\sigma^{\prime}} \delta_{\sigma\sigma^{\prime}} (\textbf{S}_{i} \cdot \bm{\sigma})_{\sigma\sigma^{\prime}} Im Tr \int_{-\infty}^{\infty} dE f_{-}(E) \sum_{\textbf{k}} \widehat{G}_{\textbf{k}}^{(0)}(E)\nonumber \\
- \frac{J^{2}}{8\pi N^{2}} \sum_{ij\textbf{q}\sigma\sigma^{\prime}} \bigg[ e^{-i\textbf{q} \cdot \big( \textbf{R}_{i} - \textbf{R}_{j} \big)} (\textbf{S}_{i} \cdot \bm{\sigma})_{\sigma^{\prime}\sigma} (\textbf{S}_{j} \cdot \bm{\sigma})_{\sigma\sigma^{\prime}} \nonumber \\
Im Tr \int_{-\infty}^{\infty} dE f_{-}(E) \sum_{\textbf{k}} \widehat{A}_{\textbf{k},\textbf{k+q}}^{\sigma\sigma^{\prime}}(E) \bigg] 
\end{align}
i.e.
\begin{align}\label{eq:hfeff}
H_{f} = -\frac{J}{2} \sum_{i\sigma} (\textbf{S}_{i} \cdot \bm{\sigma})_{\sigma\sigma} \langle n_{\sigma}^{(0)} \rangle \hspace*{4.0cm}\nonumber \\
+\frac{J^{2}}{8 N} \sum_{ij\textbf{q}\sigma\sigma^{\prime}} e^{-i\textbf{q} \cdot \big( \textbf{R}_{i} - \textbf{R}_{j} \big)} (\textbf{S}_{i} \cdot \bm{\sigma})_{\sigma^{\prime}\sigma} (\textbf{S}_{j} \cdot \bm{\sigma})_{\sigma\sigma^{\prime}} D_{\textbf{q}}^{\sigma\sigma^{\prime}}
\end{align}
where
\begin{equation}\label{eq:n0}
\langle n_{\sigma}^{(0)} \rangle = -\frac{1}{\pi} Im Tr \int_{-\infty}^{\infty} dE f_{-}(E) \frac{1}{N} \sum_{\textbf{k}} \widehat{G}_{\textbf{k}}^{(0)}(E)
\end{equation}
and 
\begin{equation}\label{eq:dqs}
D_{\textbf{q}}^{\sigma\sigma^{\prime}} = -\frac{1}{\pi} Im Tr \int_{-\infty}^{\infty} dE f_{-}(E) \frac{1}{N} \sum_{\textbf{k}} \widehat{A}_{\textbf{k},\textbf{k+q}}^{\sigma\sigma^{\prime}}(E) 
\end{equation}
If we perform spin summations on r.h.s of Eq.~(\ref{eq:hfeff}) we obtain
\begin{align}
H_{f}
= -\frac{J}{2} \sum_{i} \big( \langle n_{\uparrow}^{(0)} \rangle - \langle n_{\downarrow}^{(0)} \rangle \big) S_{i}^{z} \hspace*{6.0cm}\nonumber \\
+\frac{J^{2}}{8 N} \sum_{ij\textbf{q}} e^{-i\textbf{q} \cdot \big( \textbf{R}_{i} - \textbf{R}_{j} \big)} \hspace*{6.5cm} \nonumber \\
\bigg[ D_{\textbf{q}}^{\uparrow\downarrow}S_{i}^{-}S_{j}^{+} + D_{\textbf{q}}^{\downarrow\uparrow}S_{i}^{+}S_{j}^{-} + \big( D_{\textbf{q}}^{\uparrow\uparrow} + D_{\textbf{q}}^{\downarrow\downarrow}\big)S_{i}^{z}S_{j}^{z} \bigg] \hspace*{3.5cm}
\end{align}
where the spin operators, $S_{i}^{\zeta}$ ($\zeta$ = +,-,z ), satisfy the usual commutation relations.
The first term in the above equation is exactly zero since the free system is unpolarized. This finally yields an effective anisotropic Heisenberg- like spin Hamiltonian which can be written as follows
\begin{equation}\label{eq:effhe}
H_{f} =
- \sum_{ij} \bigg[ J_{ij}^{(1)} S_{i}^{-}S_{j}^{+} + J_{ij}^{(2)} S_{i}^{+}S_{j}^{-} + J_{ij}^{(3)} S_{i}^{z}S_{j}^{z} \bigg]
\end{equation}
where
\begin{equation}
J_{ij}^{(n)} = \frac{1}{N} \sum_{\textbf{q}} J^{(n)}(\textbf{q}) \hspace*{0.2cm}e^{-i\textbf{q} \cdot \big( \textbf{R}_{i} - \textbf{R}_{j} \big)} \hspace*{1.0cm} (\textrm{n = 1,2,3})
\end{equation}
with
\begin{eqnarray}
J^{(1)}(\textbf{q}) = -\frac{J^{2}}{8} D_{\textbf{q}}^{\uparrow\downarrow} \hspace*{0.75cm} \\
J^{(2)}(\textbf{q}) = -\frac{J^{2}}{8} D_{\textbf{q}}^{\downarrow\uparrow} \hspace*{0.75cm} \\
J^{(3)}(\textbf{q}) = -\frac{J^{2}}{8} \big(D_{\textbf{q}}^{\uparrow\uparrow} + D_{\textbf{q}}^{\downarrow\downarrow} \big)
\end{eqnarray}
are the effective exchange integrals which via $G_{\textbf{k}\sigma}$ are functionals of the conduction electron self-energy thereby getting a temperature and carrier concentration dependence. In order to obtain effective isotropic Heisenberg-like spin Hamiltonian, one can prove~\cite{sharmathesis} that $D_{\textbf{q}}^{\uparrow\downarrow}$ = $D_{\textbf{q}}^{\downarrow\uparrow}$ and $D_{\textbf{q}}^{\uparrow\uparrow} + D_{\textbf{q}}^{\downarrow\downarrow}$ = 2$D_{\textbf{q}}^{\downarrow\uparrow}$ which will result in $J^{(1)}(\textbf{q})$ = $J^{(2)}(\textbf{q})$ = $J^{(3)}(\textbf{q})$/2 = $J^{\textrm{eff}}(\textbf{q})$. 
We finally get 
\begin{equation}\label{eq:effisohe}
H_{f} =
- \sum_{ij}^{n.n} J_{ij}^{\textrm{eff}} \bigg[ \frac{1}{2} \bigg( S_{i}^{-}S_{j}^{+} + S_{i}^{+}S_{j}^{-} \bigg) + S_{i}^{z}S_{j}^{z} \bigg] \hspace*{0.25cm}
\end{equation}
\indent Now so as to deterimine the \textit{f} spin magnetization we follow along the lines of Callen~\cite{callen} that results in 
\begin{equation}\label{eq:callform}
\langle S^{z} \rangle = \frac{\big(S-\varphi\big)\big(1+\varphi\big)^{2S+1} + \big(S+1+\varphi\big)\varphi^{2S+1}}{\big(1+\varphi\big)^{2S+1} - \varphi^{2S+1}}
\end{equation}
where $\varphi(S)$ can be interpreted as average magnon number 
\begin{equation}\label{eq:magnum}
\varphi(S) = \frac{1}{N} \sum_{\textbf{q}} \frac{1}{e^{E(\textbf{q})/k_{B}T}-1}
\end{equation}
depending on S via magnon energies $E(\textbf{q})$ which can be obtained using the spin Green function~\cite{callen} and is given by
\begin{equation}\label{eq:magene}
E(\textbf{q}) = 2 \langle S^{z} \rangle \big[ J^{\textrm{eff}}(0) - {J^{\textrm{eff}}(\textbf{q})} \big] 
\end{equation}
with $J^{\textrm{eff}}(0) = J^{\textrm{eff}}(\textbf{q}=0)$. In section~\ref{sec:elesubsys}, we observed that the magnetization ($\langle S^{z} \rangle$) appears in electronic self energy. And this self energy is used to calculate the exchange integrals which along with $\langle S^{z} \rangle$ enter the magnon energies. These magnon energies in turn also appear in $\langle S^{z} \rangle$. Thus we have found a closed system of equations that can be solved self consistently for all quantities of interest, in particular those which tell us about the mutual influence of electronic and magnetic properties of the exchange coupled system of itinerant electrons and localized \textit{f} spins.\\
\indent One of the central quantities in magnetic system is the Curie temperature, $T_{c}$, which can be ascribed to the temperature for which $\langle S^{z} \rangle \rightarrow 0$. On expanding Eq.~(\ref{eq:callform}) in $\frac{1}{\varphi(S)}$ we get
\begin{equation}\label{eq:szattc}
\langle S^{z} \rangle = \frac{S(S+1)}{3\varphi(S)} + O\bigg(\frac{1}{(\varphi(S))^{2}}\bigg)
\end{equation}
\indent For $\langle S^{z} \rangle \rightarrow 0$ we have
\begin{equation}\label{eq:tayexp}
e^{E(\textbf{q})/k_{B}T} \simeq 1 + \frac{E(\textbf{q})}{k_{B}T} 
\end{equation}
\indent Using Eqs.~(\ref{eq:magnum}),~(\ref{eq:magene}),~(\ref{eq:szattc}) and~(\ref{eq:tayexp}) we obtain
\begin{equation}\label{eq:curietemp}
T_{c} = \frac{2S(S+1)}{3k_{B}} \bigg[ \frac{1}{N} \sum_{\textbf{q}} \bigg( \frac{1}{\big[ J^{\textrm{eff}}(0) - {J^{\textrm{eff}}(\textbf{q})} \big]} \bigg)_{T_{c}} \bigg]^{-1}
\end{equation}
\begin{figure}[h]
\centering
\includegraphics[height=11.5cm,width=9.0cm]{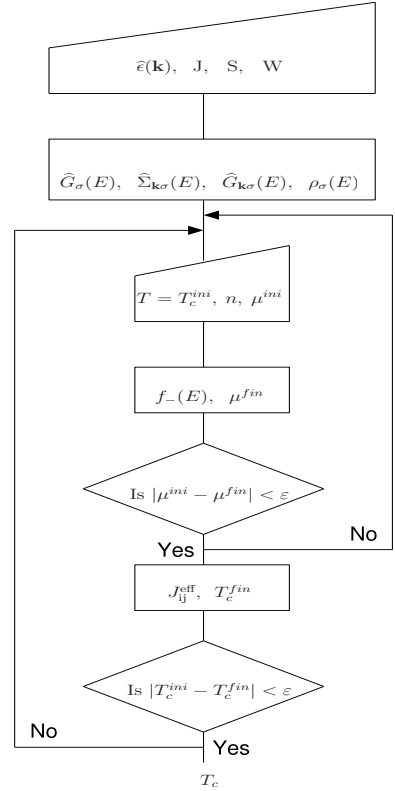}
\caption{\label{fig:flcht} Flowchart exhibiting the self consistent determination of Curie temperature, $T_{c}$. The terminologies are as explained in the text.}
\end{figure}
\indent We can evaluate Eq.~(\ref{eq:curietemp}) within a self consistent cycle as shown in Figure~\ref{fig:flcht} which can be understood as follows. In our analysis we consider a two band model ($\alpha$ and $\nu$ = 1,2) which can be generalized to a \textit{n}-band model. The single particle energies $\widehat{\epsilon}(\textbf{k})$ is then represented in a 2 x 2 matrix where the diagonal and non-diagonal terms are considered to have the following form. $\epsilon^{11}(\textbf{k}) = -\frac{\textrm{W}}{6} (cos(k_{x}a)+cos(k_{y}a)+cos(k_{z}a))$ ,  $\epsilon^{12}(\textbf{k}) = \epsilon^{21}(\textbf{k}) = V$ (local hybridization, LH) or $\epsilon^{12}(\textbf{k}) = \epsilon^{21}(\textbf{k}) = V\epsilon^{11}(\textbf{k})$ (non-local hybridization, NLH) and $\epsilon^{22}(\textbf{k}) = E_{0} + \epsilon^{11}(\textbf{k})$. Along with the single particle energies, the intra-atomic exchange J, quantum spin number S and bandwidth W act as input parameters in order to evaluate the free propagator, self-energy and full propagator. Then for a particular band occupation, $n$, and an initial temperature $T_{c}^{ini}$, the Fermi edge, $\mu$ which yields the correct value of $n$, is determined self consistently. Thus, fixing upon the Fermi edge one evaluates the temperature dependent exchange integrals which gives the $T_{c}$ through Eq.~(\ref{eq:curietemp}). If the obtained temperature is within convergence limit, $\varepsilon$, then it is the resulting $T_{c}$ for particular $J$ and $n$. \\

\section{Results $\&$ Discussion}
\label{sec:resdis}

\indent On optimizing the numerical factor~\cite{appn}, we evaluate the Curie temperature, i.e. Eq.~(\ref{eq:curietemp}), for various configurations of model parameters (J, n and V). We consider the spin quantum number S = $\frac{7}{2}$ (localized moment of Eu$^{2+}$, Gd$^{3+}$) and the center of gravity of the second band shifted by amount, $E_{0}$ = 0.25 eV. First we try to reproduce the single band result for different values of band occupation as was previously obtained but using another electronic self-energy~\cite{santos2}. This will give us some confidence on the working of the algorithm. \\
\begin{figure}[!h]
\centering
\vspace*{0.3cm}
\includegraphics[height=7.0cm,width=8.5cm]{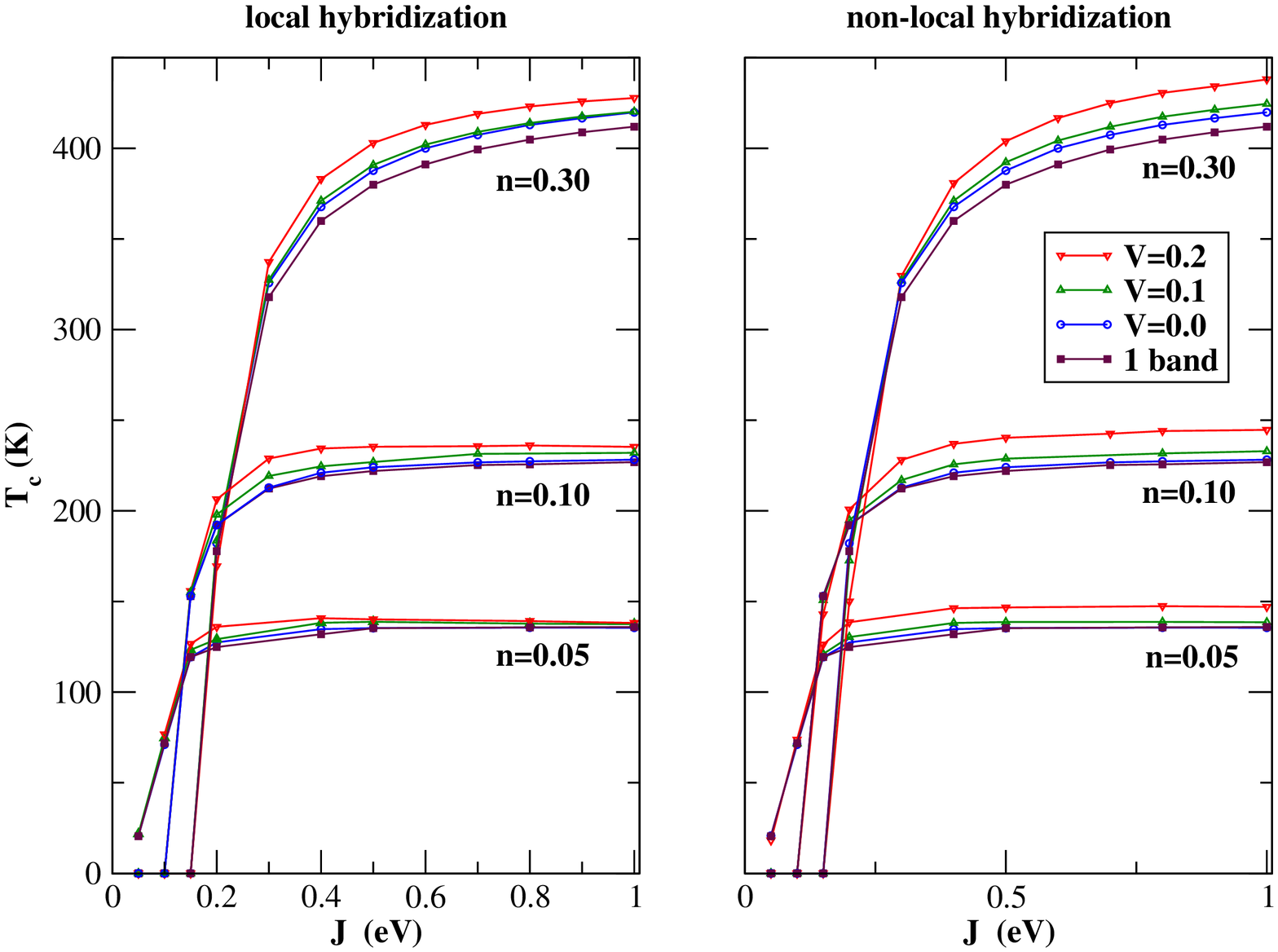}
\vspace*{0.5cm}
\caption{\label{fig:tcvsjsf}(Color online) The dependence of Curie temperature ($T_{c}$) on intra-atomic exchange ($J$) for different values of band occupation. The exhibited results are for single and two band KLM with local (left panel) and non-local hybridization (right panel) on a simple cubic lattice.}
\end{figure}
\indent Figure~\ref{fig:tcvsjsf} shows the dependence of $T_{c}$ on the strength of intra-atomic exchange, $J$, for single and two band KLM. The left panel describes the results with local hybridization (LH) while the right panel for non-local hybridization (NLH) on a simple cubic lattice. The bandwidth of both the bands, W, is taken to be 1.0 eV. The calculations are carried out for different values of band occupation, $n$, and hybridization, $V$.\\
\indent Though both the graphs look quite similar but there are marked differences especially in the limit of strong coupling and low band occupation. We first discuss the general behavior. It is observed that initially the $T_{c}$ rises sharply with increasing $J$. For weak coupling and small band occupation (n=0.05), the usual RKKY mechanism is observed. However for higher band occupation, $J$ is observed to exceed a critical value in order to allow ferromagnetism. Furthermore, with increasing $J$ the critical temperature is observed to be deviating more and more from the RKKY behavior (i.e., long- range order) and finally reaches a saturation. The calculations done within single band model are comparable with previous calculations~\cite{santos2} obtained using a different self-energy. \\
\indent As shown in Figure~\ref{fig:tcvsjsf}, the results for $T_{c}$ in case of two unhybridized bands ($V$=0.0) are in comparison with that of the one band situation for low band occupation due to similar low energy paramagnetic density of states at the Fermi edge. And with increasing band occupation the results in both the situation (local and non-local hybridization) differ drastically. The values of Curie temperatures are found to be higher and increasing with increasing hybridization strength and band occupation for two hybridized bands as compared to the unhybridized or one band model. It can be understood as follows. \\
\begin{figure}[!h]
\centering
\includegraphics[height=7.0cm,width=8.0cm]{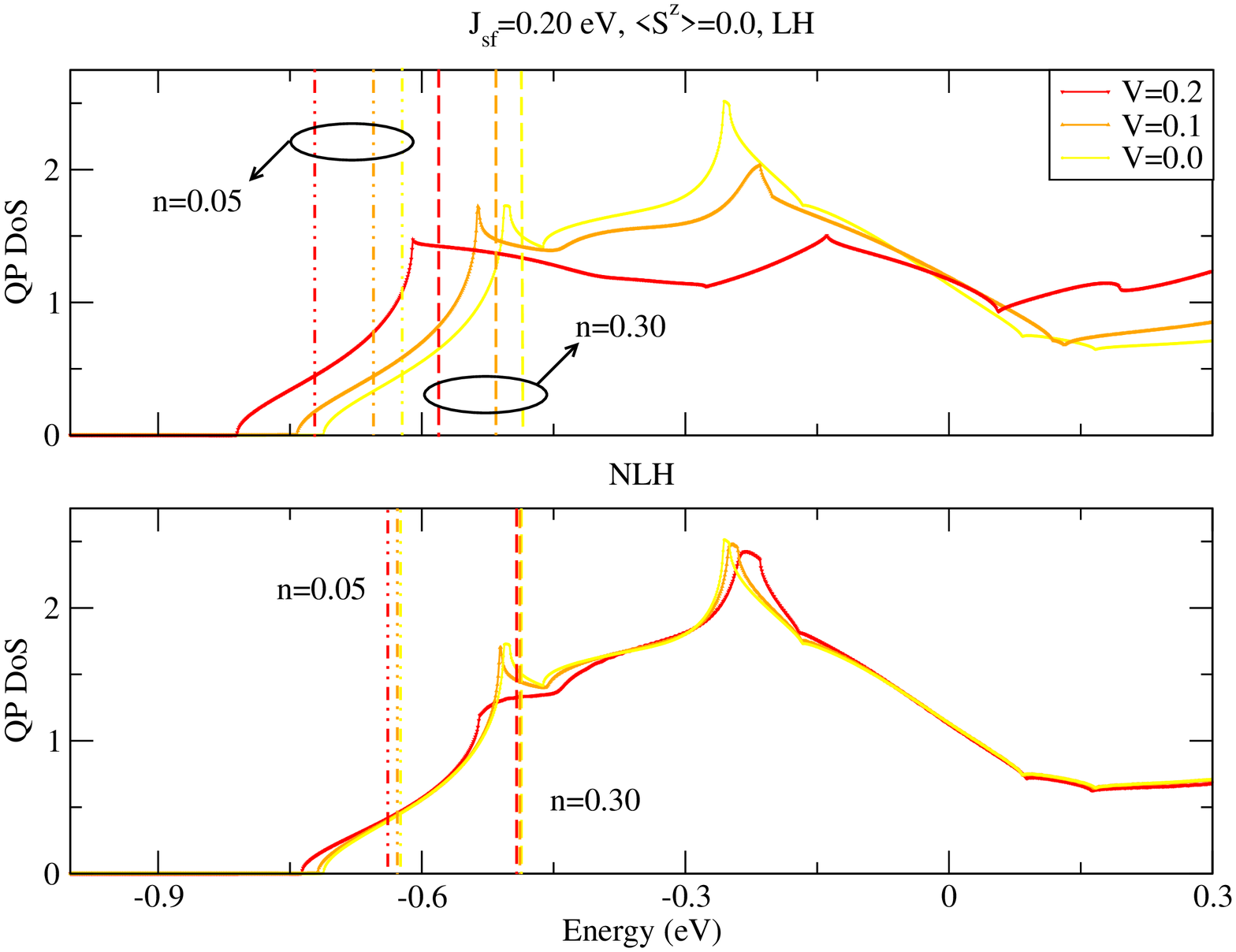}
\caption{\label{fig:dosj02}(Color online) Lower edge of paramagnetic density of states of two band KLM with LH (upper panel) and NLH (lower panel) on a sc lattice for $J$=0.20 eV. The curves and vertical lines in yellow, orange and red are for $V$=0.0, 0.1 and 0.2 respectively. The vertical lines denotes the Fermi edge with dashed-dotted and dashed lines for a band occupation, $n$, of 0.05 and 0.30 respectively. }
\end{figure}
\indent Figure~\ref{fig:dosj02} shows the low energy window of the paramagnetic density of states of two band KLM with local and non-local hybridization shown in upper and lower panel respectively. The calculations are performed for $J$=0.20 eV. The curves and vertical lines in yellow, orange and red are for $V$=0.0, 0.1 and 0.2. The vertical lines signifies the Fermi edge with dashed-dotted and dashed lines for a band occupation of 0.05 and 0.30 respectively. The lines within the circle represent the Fermi edge for different values of hybridization but for the same value of band occupation.\\
\indent It is to be noted that in general the shape of the density of states is dependent on the band occupation which is a consequence of electronic correlation effects. But the self-energy which we consider in our calculations is independent of band occupation or rather well defined only in the limit of low band occupation. Since we restrict ourselves to this limit, thus we have density of states dependent only on the strength of hybridization. The band filling is determined by the placement of Fermi edge which is obtained self-consistently in our calculations. In order to improve over the restricted limit, we can consider the band occupation dependent self-energy as given in Ref.~\onlinecite{nolisa2}. But this is not the aim of the present paper.\\
\indent We observe that for low band occupation, the density of states at the Fermi edge are slightly different for different values of hybridization. But with increasing band occupation, the density of states at the Fermi edge changes abruptly for local as well as non-local hybridization. \\
\begin{figure}[!h]
\centering
\includegraphics[height=7.0cm,width=8.5cm]{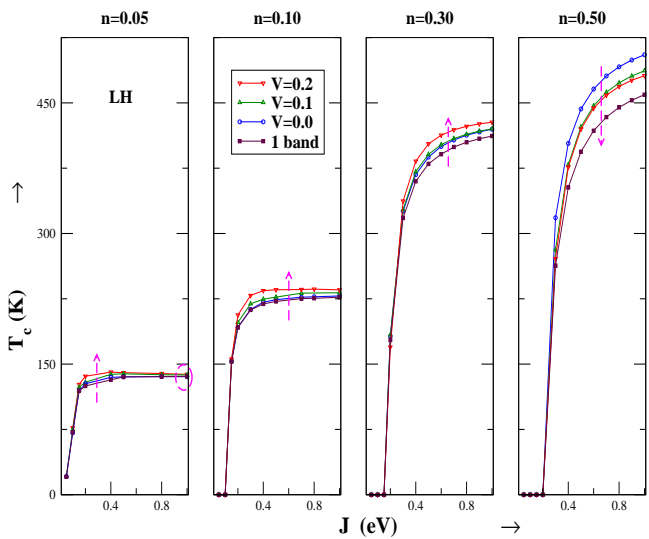}
\caption{\label{fig:tcvjvarn}(Color online) The same as in Figure~\ref{fig:tcvsjsf} but only for LH and an additional result for band occupation of n=0.50.}
\end{figure}
\indent Though for high band occupation the values of $T_{c}$ in case of two hybridized bands are higher as compared to unhybridized or one band situation but for low band occupation and strong coupling limit, an interesting feature is observed as shown (encircled) in Figure~\ref{fig:tcvjvarn}. Since it is only observed in case of two locally hybridized bands so we do not consider the case of non-local hybridization. It is noted that in the limit of strong coupling the $T_{c}$ starts decreasing for two hybridized band system as compared to two unhybridized band model. Even though increasing hybridization increases the bandwidth and therefore the kinetic energy of the itinerant electrons but it is only effective for higher band occupation. In the regime of low band occupation and strong coupling the short range order due to strong intra-atomic exchange and local hybridization decreases the kinetic energy leading to localization of the carrier. This results in decrease of the paramagnetic density of states at the Fermi edge thereby reducing the $T_{c}$ to the value of the one band case which is encircled in the left-most panel in Figure~\ref{fig:tcvjvarn}. But as mentioned earlier, upon increasing the band occupation (moving to the right panels) we observe that the $T_{c}$ increases with increase in hybridization (bandwidth~\cite{santos2}) and also in the limit of strong coupling due to the presence of more delocalized electrons. It can be again understood within the picture of the density of states. \\
\begin{figure}[!h]
\centering
\includegraphics[height=4.5cm,width=7.0cm]{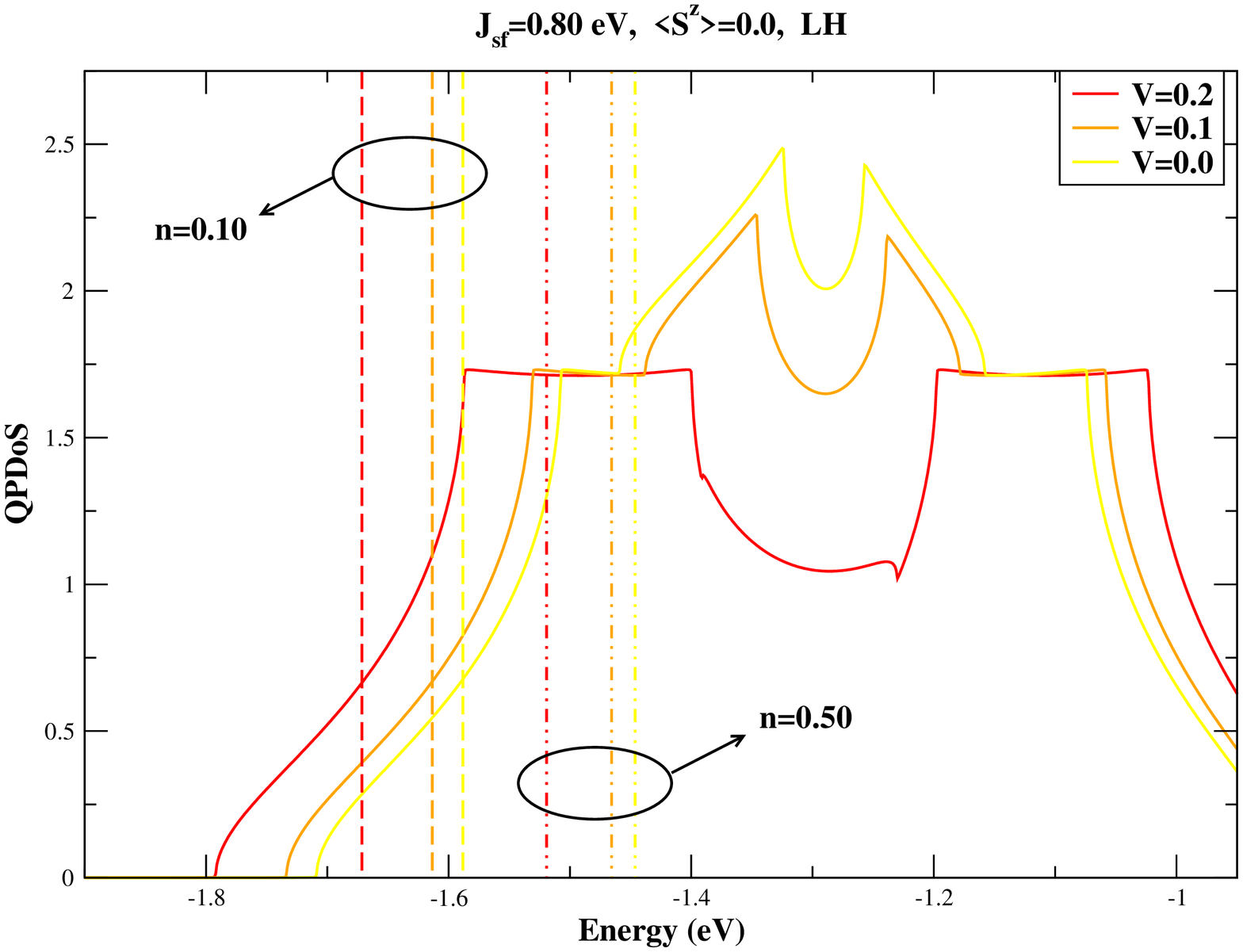}
\caption{\label{fig:doslhj08}(Color online) Lower edge of paramagnetic density of states for two locally hybridized band KLM on a sc lattice for $J$=0.80 eV. The curves and vertical lines in yellow, orange and red are for $V$=0.0, 0.1 and 0.2 respectively. The vertical lines denote the Fermi edge with dashed-dotted and dashed lines for a band occupation, $n$, of 0.10 and 0.50 respectively.}
\end{figure}
\indent Figure~\ref{fig:doslhj08} shows the low energy spectrum of paramagnetic density of states for two locally hybridized band KLM on a simple cubic lattice for $J$=0.80 eV. The curves and vertical lines in yellow, orange and red are for $V$=0.0, 0.1 and 0.2. The vertical lines denote the Fermi edge with dashed-dotted and dashed lines for a band occupation of 0.10 and 0.50 respectively. As observed for $n$=0.10 that the density of states at the Fermi edge are slightly different giving rise to marginal difference in $T_{c}$ with increase in hybridization. But with increasing band occupation ($n=0.50$), the density of states at the Fermi edge differ drastically giving rise to quite different Curie temperatures. \\
\begin{figure}[!h]
\centering
\includegraphics[height=4.5cm,width=7.0cm]{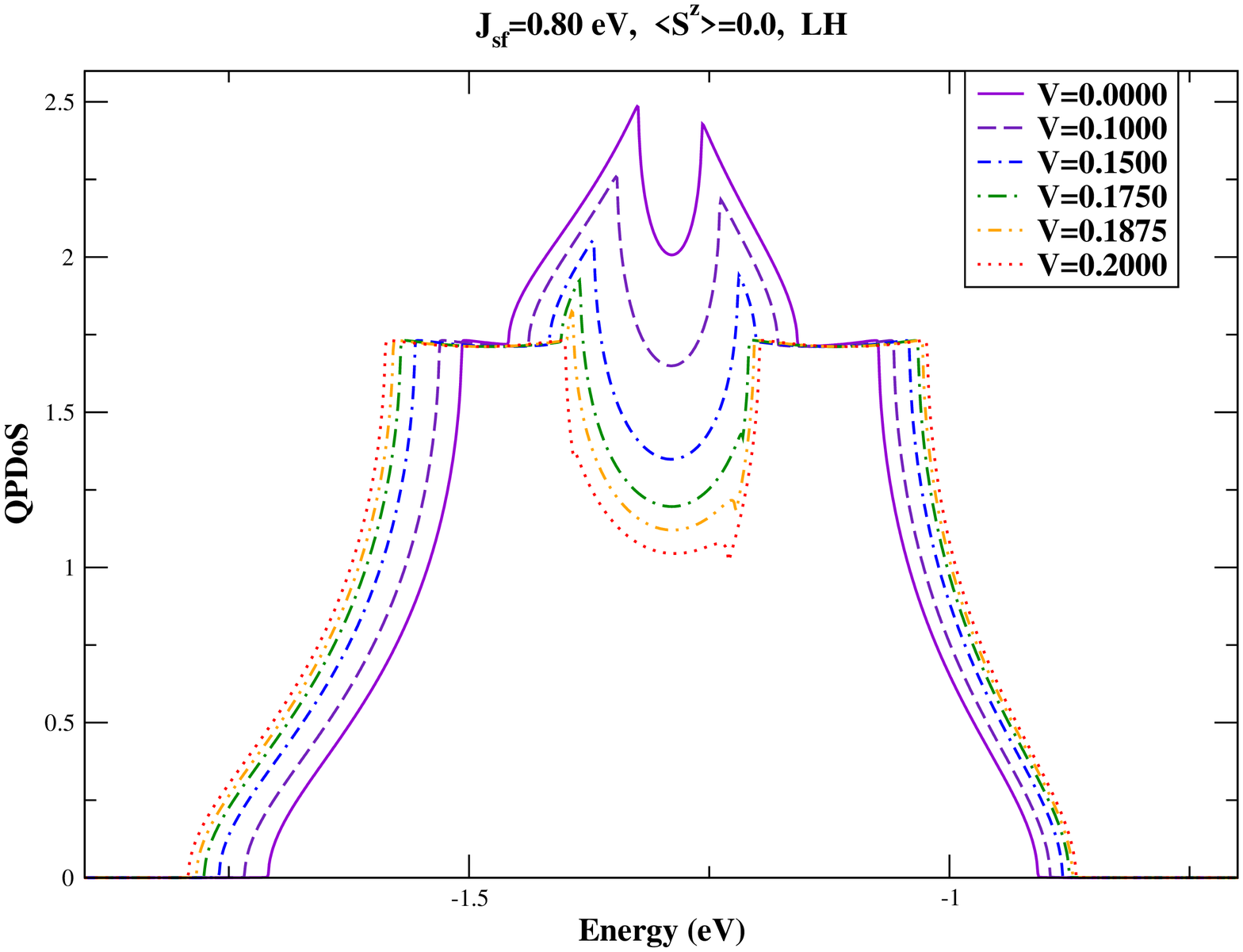}
\caption{\label{fig:doslhj08n}(Color online) Lower edge of paramagnetic density of states for two locally hybridized band KLM on a sc lattice for $J$=0.80 eV and for different values of hybridization.}
\end{figure}
\indent Another feature which we observe in the strong coupling limit is that the density of states tend to seperate out with increasing value of hybridization as shown in Figure~\ref{fig:doslhj08}. But the change from $V$=0.1 to $V$=0.2 is quite sudden. In order to have a close look at it, we plot the paramagnetic density of states for $J$=0.80 eV and intermediate values of hybridization from $V$=0.1 to $V$=0.2 as shown in Figure~\ref{fig:doslhj08n}. It is noted that there is an increase in bandwidth with increasing hybridization. \\
\indent Further interesting characteristic is observed for band occupation of $n$=0.50. It is seen that the trend of higher and increasing value of $T_{c}$ for two locally hybridized band model is reversed as shown in the right most panel of Figure~\ref{fig:tcvjvarn}. As the Fermi edge keeps on moving to higher energies with an increase in band occupation, the paramagnetic density of states at the Fermi edge keeps on changing. This results in an observed change in the $T_{c}$. 
A similar pattern of decrease in the value of Curie temperature with increase in band occupation and strength of hybridization is also observed in case of non-local hybridization. The explanation lies similar to what we have explained earlier in case of local hybridization.\\ 
\begin{figure}[!h]
\centering
\vspace*{0.3cm}
\includegraphics[height=7.0cm,width=8.5cm]{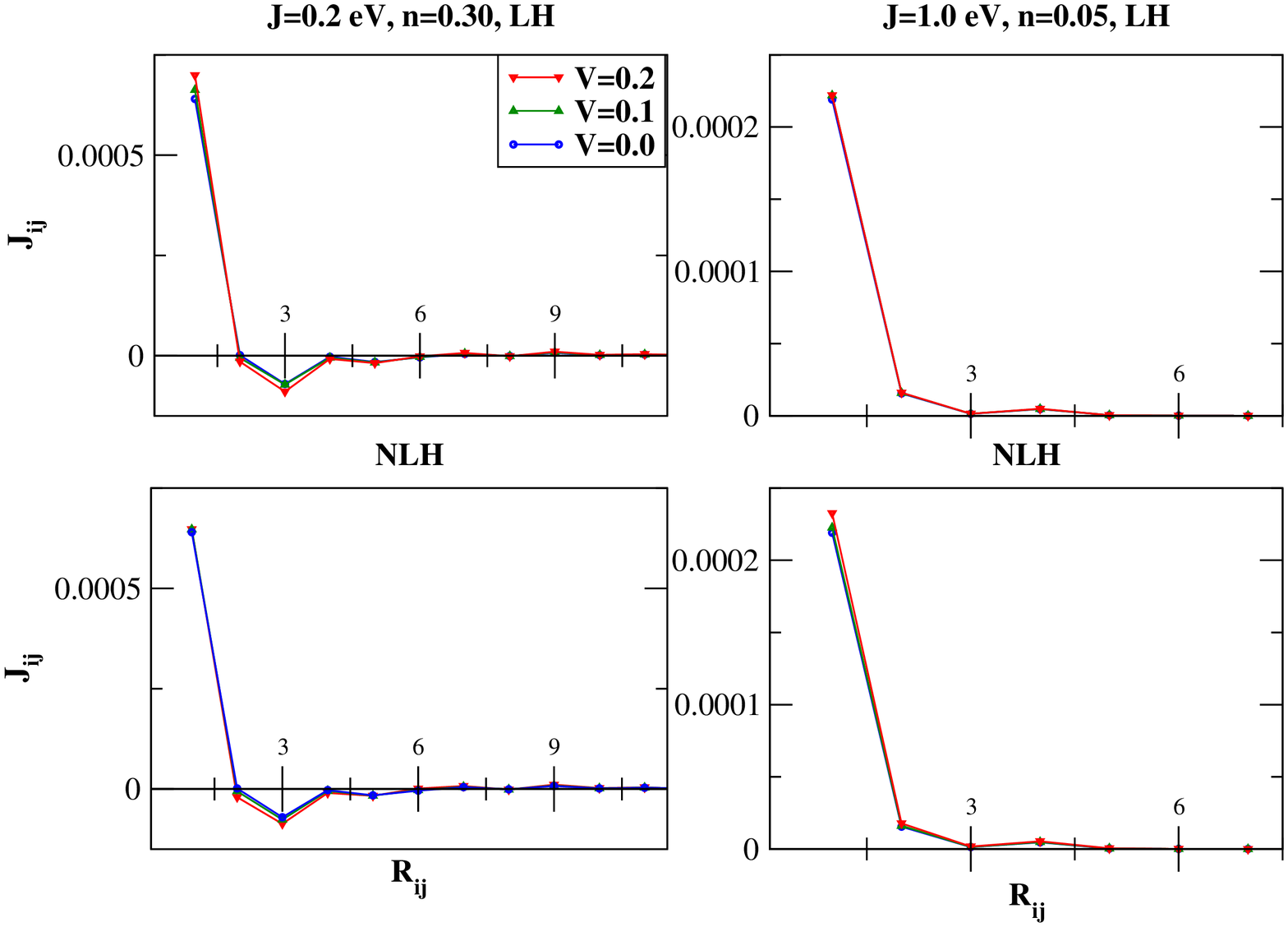}
\caption{\label{fig:exchintlnlh}(Color online) The indirect exchange integrals shown as a function of distance for local (upper panel) and non-local hybridization (lower panel). The shown results are for two different values of intra-atomic exchange and band occupation but for three different values of hybridization. }
\end{figure}
\indent One the other hand since $T_{c}$ is directly related to effective exchange integrals, $J_{ij}^{\textrm{eff}}$, it is also interesting to notice the behavior of these exchange integrals for different values of $n$, $J$ and $V$ and for local as well as non-local hybridization. Figure~\ref{fig:exchintlnlh} shows the dependence of indirect effective exchange integrals on two different parameter configuration of $J$, $n$ and for three different values of hybridization. We observe the long- range RKKY kind of oscillations~\cite{rkky} for weak coupling in case of local and non-local hybridization. In case of strong coupling, the local short range order is more strong. In that case, the exchange integrals get converged very quickly within a short distance. \\

%


\section{Summary $\&$ Conclusion}
\label{sec:sumcon}

\indent In this paper, we studied the magnetic properties of the multi-band Kondo lattice model Hamiltonian which describes the intra-atomic exchange interaction between itinerant electrons and localized spins on a periodic lattice. In section~\ref{sec:elesubsys}, we considered only the electronic part of the system. Using an earlier proposed ansatz for multi-band self-energy~\cite{sharmagdn}, we evaluated the electronic properties of interest like the density of states and band occupation number. In section~\ref{sec:magsubsys}, we developed the modified RKKY theory~\cite{nolrexmat} wherein we integrated out the charge degrees of freedom of the itinerant electrons thereby mapping the multi-band Kondo lattice model Hamiltonian onto an effective Heisenberg- like spin Hamiltonian. In section~\ref{sec:resdis}, using this procedure we determined the magnetic properties of the system like Curie temperature (within Random Phase Approximation) for various values of system parameters while calculating the chemical potential and magnetization within a self consistent scheme. \\
\indent We found that the $T_{c}$ as a function of intra-atomic exchange J for a two band KLM remains qualitatively the same for local as well as non-local hybridization between both the bands except for the limit of low band occupation and strong coupling. For higher band occupation and increase in strenght of coupling as well as hybridization we find that $T_{c}$ increases until the band occupation of n=0.5 from where the trend is reversed. All these can be explained using the paramagnetic density of states and its behaviour at the Fermi edge. It is mainly due to the interplay between kinetic and potential energy. In case of strong coupling, the $T_{c}$ is oscillating in its dependence on the band occupation. \\
\indent Such an analysis can be very useful in order to understand the physical properties of real materials described within the multi-band models like the manganites. These materials are known to have a strong intra- atomic exchange coupling behavior. Or the analysis can be equally handful for the rare- earth metals which are known to be described within the weak or inter- mediate intra- atomic coupling regime. It would be equally encouraging to carry out the similar investigation for two different bandwidths of both the bands since the correlation effects scale as $\frac{J}{W}$ where $W$ being the bandwidth. \\
\indent We would also like to apply~\cite{shagdnmag} the multi- band modified RKKY theory in order to understand the basic mechanism behind the observed ferromagnetism in GdN~\cite{li,granville}.


\section{Acknowledgement}
\label{sec:ack}

One of the authors (A.S.) benefitted from valuable discussions on the numerical calculations with S\"oren Henning and Jochen Kienert.


\begin{thebibliography}{99}
\bibitem{kondo} J.Kondo, Prog. Theor. Phys. \textbf{32}, 37 (1964).
\bibitem{furu} N.Furukawa, J. Phys. Soc. Jpn \textbf{63}, 3214 (1994).
\bibitem{dago} E.Dagotto, S.Yunoki, A.L.Malvezzi, A.Moreo, J.Hu, S.Capponi, D.Poilblanc, and N.Furukawa, Phys. Rev. B \textbf{58}, 6414 (1998).
\bibitem{kasu} T.Kasuya, Prog. Theo. Phys. \textbf{16}, 45 (1956).
\bibitem{yosi} K.Yosida, Phys. Rev. \textbf{106}, 893 (1957).
\bibitem{nol79} W.Nolting, Phys. Status Sol. B \textbf{96}, 11 (1979).
\bibitem{nagaev} E.L.Nagaev, \textit{Physics of magnetic semiconductors}, Mir, Moscow, (1983).
\bibitem{wachter} P.Wachter, \textit{Handbook on the Physics and Chemistry of Rare Earth}, \textbf{Chapter 19}, North Holland Publishing Company, Amsterdam, (1979).
\bibitem{mauger} A.Mauger and C.Godart, Physics Reports (Review Section of Physics Letters) \textbf{141}, 51-176 , North Holland Publishing Company, Amsterdam, (1986).
\bibitem{mueller} W.M\"uller and W.Nolting, Phys. Rev. B \textbf{69}, 155425 (2004).
\bibitem{dietl} T. Dietl and H. Ohno, \textit{Ferromagnetic III-V and II-VI Semiconductors}, MRS Bulletin, p. 714 (2003).
\bibitem{tang} G.Tang and W.Nolting, Phys. Rev. B \textbf{75}, 024426 (2007).
\bibitem{singh} A.Singh, S.K.Das, A.Sharma and W.Nolting, J. Phys.:Condens. Matter \textbf{19}, 236213 (2007).
\bibitem{wolf} S.A.Wolf, D.D.Awschalom, R.A.Buhrman, J.M.Daughton, S.von Moln$\acute{a}$r, M.L.Roukes, A.Y.Chtchelkanova and D.M.Treger, Science \textbf{294}, 1488 (2001).
\bibitem{zutic} I.$\check{Z}$uti$\acute{c}$, J.Fabian and S.Das Sarma, Rev.Mod.Phys. \textbf{76}, 323 (2004).
\bibitem{rexeyenol} S.Rex, V.Eyert, and W.Nolting, J. Magn. Magn. Mater. \textbf{192}, 529 (1999).
\bibitem{santos} C.Santos, W.Nolting, and V.Eyert, Phys. Rev. B \textbf{69}, 214412 (2004).
\bibitem{lizhang} D.Li, J.Zhang, P.A.Dowben, and M.Onellion, Phys. Rev. B \textbf{45}, 7272 (1992).
\bibitem{dondownol} \textit{Magnetism and Electronic correlations in local moment systems : Rare earth elements and compounds} edited by M.Donath, P.Dowben and W.Nolting, World Scientific, Singapore (1998).
\bibitem{jin} S.Jin, T.H.Tiefel, M. McCormack, R.A.Fastnacht, R.Ramesh, and L.H.Chen, Science \textbf{264}, 413 (1994).
\bibitem{ramirez} A.P.Ramirez, J.Phys.:Condens. Matter \textbf{9}, 8171 (1997).
\bibitem{doublexchange} C.Zener, Phys. Rev. \textbf{81}, 440 (1951); P.W.Anderson and H.Hasegawa, Phys. Rev. \textbf{100}, 675 (1955); P.G.de Gennes, Phys. Rev. \textbf{118}, 141 (1960); K.Kubo and N.Ohata, J. Phys. Soc. Jpn. \textbf{33}, 21 (1972); N.Ohata, J. Phys. Soc. Jpn. \textbf{34}, 343 (1973).
\bibitem{stiernol} M.Stier and W.Nolting, Phys. Rev. B \textbf{75}, 144409 (2007).
\bibitem{loeh} H.von L\"ohneysen in Ref.~\onlinecite{dondownol}.
\bibitem{doniach} S.Doniach, Physica B $\&$ C \textbf{91}, 231 (1977).
\bibitem{popescu} F.Popescu, PhD thesis, Florida State University, \textit{Multiband models for colossal magnetoresistance materials and diluted magnetic semiconductors} (2007).
\bibitem{nagao} H.Nagao, Int. Journ. Quan. Chem. \textbf{100}, 867 (2004).
\bibitem{hub} J.Hubbard, Proc. Roy. Soc. A \textbf{276}, 238 (1963).
\bibitem{penc} K.Penc, H.Shiba, F.Mila, and T.Tsukagoshi, Phys. Rev. B \textbf{54}, 4056 (1996).
\bibitem{nolbor} W.Nolting, and W.Borgiel, Phys. Rev. B \textbf{39}, 6962 (1989).
\bibitem{wang} Y.-Q.Wang, H.Q.Lin and J.E.Gubernatis, Commun. Comput. Phys. \textbf{1}, 575 (2006) 
\bibitem{pam} P.W.Anderson, Phys. Rev. \textbf{124}, 41 (1961).
\bibitem{rkky} M.A.Ruderman, and C.Kittel, Phys. Rev. \textbf{96}, 99 (1954); T.Kasuya, Prog. Theor. Phys. \textbf{16}, 45 (1956); K.Yosida, Phys. Rev. \textbf{106}, 893 (1957).
\bibitem{meyer} D.Meyer, W. Nolting, G.G.Reddy, A.Ramakanth, Phys. Status Solidi B \textbf{208}, 473 (1999).
\bibitem{sharmagdn} A.Sharma and W.Nolting, J. Phys.: Condens. Matter \textbf{18}, 7337 (2006).
\bibitem{nolrexmat} W.Nolting, S.Rex, and S.Mathi Jaya, J.Phys.:Condens. Matter \textbf{9}, 1301 (1997).
\bibitem{zubarev} D.N.Zubarev, Usp. Fiz. Nauk \textbf{71}, 71 (1960); [English transl.: Soviet Phys. Usp. \textbf{3}, 320 (1960)].
\bibitem{sharmaeus} A.Sharma and W.Nolting, Phys. Stat. Solidi B \textbf{243}, 641 (2006). In this article, the case of carriers in completely filled bands was considered. But it can be proved that the case of carriers in completely empty bands, as considered in present manuscript, is identical to that of the fully occupied bands.
\bibitem{sharmathesis} A.Sharma, PhD thesis, Humboldt-Universit\"at zu Berlin, \textit{Electronic correlation and magnetism in multi-band Kondo lattice model} (2007).
\bibitem{callen} H.B.Callen, Phys. Rev. \textbf{130}, 890 (1963).
\bibitem{appn} The small imaginary part has to be numerically optimized for low band occupation and coupling strength.
\bibitem{santos2} C.Santos and W.Nolting, Phys. Rev. B \textbf{65}, 144419 (2002).
\bibitem{nolisa2} W.Nolting, G.G.Reddy, A.Ramakanth, D.Meyer, and J.Kienert, Phys. Rev. B \textbf{67}, 024426 (2003). 
\bibitem{shagdnmag} A.Sharma and W.Nolting, \textit{to be published}.
\bibitem{li} D.X.Li, Y.Haga, H.Shida, T.Suzuki, Y.S.Kwon, and G.Kido, J. Phys.: Condens. Matter \textbf{9}, 10777 (1997).
\bibitem{granville} S.Granville, B.J.Ruck, F.Budde, A.Koo, D.J.Pringle, F.Kuchler, A.R.H.Preston, D.H.Housden, N.Lund, A.Bittar, G.V.M.Williams, and H.J.Trodahl, Phys. Rev. B \textbf{73}, 235335 (2006).

\end{thebibliography}
\end{document}